\documentclass[showpacs,amsmath,amssymb,twocolumn,superscriptaddress,notitlepage,preprintnumbers,pra]{revtex4-1}
\usepackage[dvips]{graphicx}
\usepackage{amsmath,amssymb,amsthm,mathrsfs,amsfonts,dsfont}
\usepackage{subfigure, epsfig}
\usepackage{braket}
\usepackage{bm}
\usepackage{enumerate}
\usepackage{color,xcolor}
\usepackage{qcircuit}
\usepackage{graphicx}
\usepackage{algorithm}
\usepackage{algpseudocode}
\usepackage{algpseudocode}
\usepackage{hyperref}
\hypersetup{colorlinks=true, linkcolor=blue, citecolor=blue, urlcolor=black }
\usepackage{comment}
\usepackage[normalem]{ulem}

\newcommand{\tr}{\operatorname{Tr}} 

\newcommand{\RE}{\operatorname{Re}}
\newcommand{\IM}{\operatorname{Im}}

\newcommand{\Var}{\mathrm{Var}}

\usepackage{varioref}
\labelformat{equation}{(#1)}

\labelformat{section}{#1}

\labelformat{figure}{#1}

\labelformat{proposition}{#1}

\labelformat{lemma}{#1}

\labelformat{theorem}{#1}

\labelformat{observation}{#1}

\labelformat{definition}{#1}

\labelformat{problem}{#1}

\labelformat{algorithm}{#1}

\labelformat{corollary}{#1}

\labelformat{statement}{#1}

\newtheorem{proposition}{Proposition}

\newtheorem{corollary}{Corollary}

 \newcommand{\sun}[1]{\textcolor{black}{#1}}
\begin{document}

\title{ Randomised composite linear-combination-of-unitaries: \\its role in quantum simulation and observable estimation}

\date{\today}

% --------------------  ABSTRACT  --------------------

% ------------  AUTHORS AND AFFILIATIONS ----------
\author{Jinzhao Sun}
\email{jinzhao.sun.phys@gmail.com}
\affiliation{School of Physical and Chemical Sciences, Queen Mary University of London, London E1 4NS, UK }

\author{Pei Zeng}
\email{peizeng@uchicago.edu}
\affiliation{ Pritzker School of Molecular Engineering, The University of Chicago, Illinois 60637, USA }

\begin{abstract}

Randomisation is widely used in quantum algorithms to reduce the number of quantum gates and ancillary qubits required. A range of randomised algorithms, including eigenstate property estimation by spectral filters, Hamiltonian simulation, and perturbative quantum simulation, though motivated and designed for different applications, share common features in the use of unitary decomposition and Hadamard-test-based implementation. In this work, we start by analysing the role of randomised linear-combination-of-unitaries (LCU) in quantum simulations, and present several quantum circuits that realise the randomised composite LCU. A caveat of randomisation, however, is that the resulting state cannot be deterministically prepared, which often takes an unphysical form $U \rho V^\dagger$ with unitaries $U$ and $V$. Therefore, randomised LCU algorithms are typically restricted to only estimating the expectation value of a single Pauli operator. To address this, we introduce a quantum instrument that can realise a non-completely-positive map, whose feature of frequent measurement and reset on the ancilla makes it particularly suitable in the fault-tolerant regime. We then show how to construct an unbiased estimator of the effective (unphysical) state $U \rho V^\dagger$ and its generalisation. Moreover, we demonstrate how to effectively realise the state prepared by applying an operator that admits a composite LCU form. Our results reveal a natural connection between randomised LCU algorithms and shadow tomography, thereby allowing simultaneous estimation of many observables efficiently. As a concrete example, we construct the estimators and present the simulation complexity for three use cases of randomised LCU in Hamiltonian simulation and eigenstate preparation tasks.

\end{abstract}

\maketitle

\section{Introduction}
 % the core idea of random sampling is to avoid preparing a full quantum state or implementing a full LCU operation, and instead focus on estimating the expectation value of observables.

Reducing the computational resources (qubits and gate count) necessary for efficiently implementing quantum algorithms is desirable for noisy intermediate-scale quantum (NISQ) devices and early fault-tolerant quantum computing (FTQC) platforms~\cite{katabarwa2024early,preskill2025beyond}.
While quantum signal processing techniques based on block encoding~\cite{low2019hamiltonian} offer optimal query complexity for tasks such as Hamiltonian simulation~\cite{low2017optimal} and ground-state preparation~\cite{lin2020near,gilyen2019quantum}, they often incur significant overhead; thereby motivating alternative approaches that avoid block encoding and multi-qubit controls. Among those, random-sampling methods~\cite{lin2022heisenberg,zeng2021universal,wang2023quantum,zhang2022computing,yang2021accelerated,wan2021randomized,lin2021heisenberglimited,lu2021algorithms,zhang2022computing,huo2021shallow,wang2023quantum,ding2024quantum,he2022quantum,wang2023faster,wang2024qubit,sun2023probing,kiss2025early,faehrmann2022randomizing} have emerged as an attractive approach that preserves favourable scaling, making them suitable for early FTQC. For instance, Trotter-LCU types of algorithms~\cite{childs2012Hamiltonian,faehrmann2021randomizing,faehrmann2024short,miessen2023quantum,childs2021theory,ikeda2024measuring} for Hamiltonian simulation can eliminate Trotter error and enable high-precision dynamics~\cite{zeng2022simple,yang2021accelerated,wan2021randomized}, whereas random-sampling spectral filters have shown efficient ground-state energy and property estimation~\cite{yang2021accelerated,wan2021randomized,lin2021heisenberglimited,zeng2021universal,lu2021algorithms,zhang2022computing,huo2021shallow,wang2023quantum,ding2024quantum,he2022quantum,wang2023faster,wang2024qubit,sun2023probing,ding2023even,ding2023robust}. Concurrently, there have been efforts to reduce qubit requirements for simulating large-scale quantum dynamics on smaller devices~\cite{yuan2020quantum,peng2020simulating,schuhmacher2025hybrid}, including perturbative quantum simulation (PQS)~\cite{sun2021perturbative} and circuit knitting~\cite{harrow2025optimal,peng2020simulating}.

\begin{figure*}[t!]
\centering
\includegraphics[width=1\textwidth]{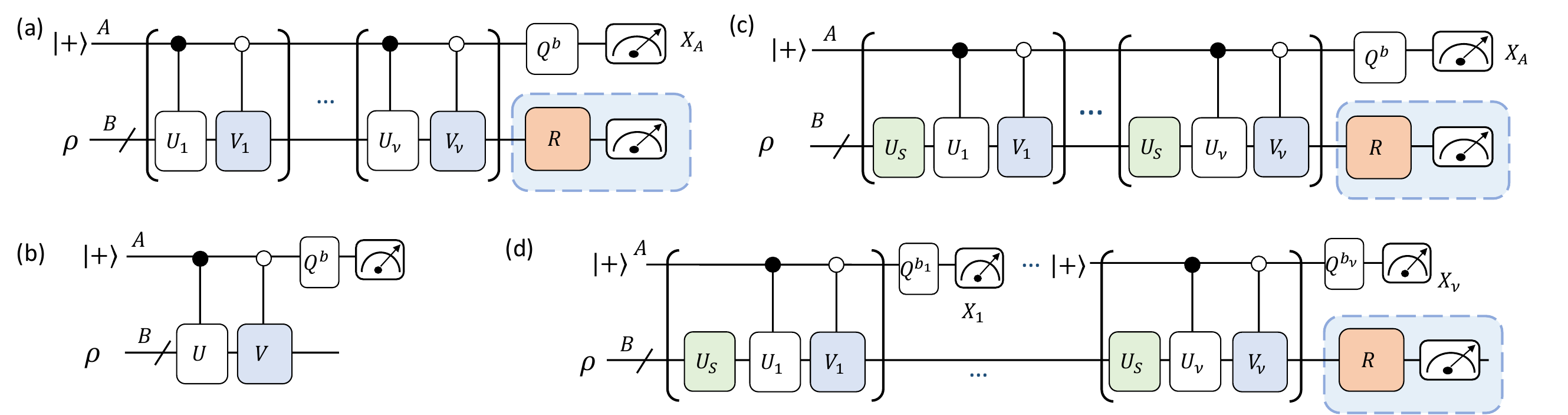}
\caption{ (a) The quantum circuit for generating an unbiased estimator for the non-physical state $ \prod_{k = 1}^{\nu} U_k \rho \prod_{k = 1}^{\nu}  V_{k}^{\dagger} $ and observable estimation by classical shadows. The unitary $U_{k}$ and $V_k$ for $k = 1,..., \nu$ are randomly sampled from the distribution by \autoref{eq:product_LCU}.  Here, $b = \{0, 1\}$ is generated uniformly at random which determines whether the inverted phase gate  ($Q = S^{\dagger}$) is applied non-trivially. The shadow estimation is applied in the end. The dotted box area in (a,c,d) represents the process of shadow estimation.  $R$ represent the random Clifford operations chosen from a tomographically complete set, as used in the conventional classical shadow method~\cite{huang2020predicting}. 
(b) The standard Hadamard test circuit is a special case of (a), corresponding to a single segment $\nu = 1$. 
(c)  The circuit for generating unbiased estimator for $ \prod_{k = 1}^{\nu} U_k U_S \rho \prod_{k = 1}^{\nu} U_S^{\dagger} V_{k}^{\dagger} $, which can be used to estimate the properties of real-time evolved states.  A common unitary $U_S$ appears in all segments, as is often the case in Trotterised evolution.
(d)  The ancilla is initialised and reset in each segment.
In each of the segments, we add the ancilla initialised in $\ket{+}$, apply the controlled-unitaries followed by an inverted quantum phase gate and perform the measurement on the $X$ basis with measurement outcome $a_k = \{0, 1\}$.
Here, $b_k = \{0, 1\}$ is generated uniformly at random.
The measurement outcome on the ancilla and the circuit setting is recorded as $(a_k, b_k )$. To realise the composite LCU, the circuit instance is fixed with either $b_k = 0$ or $b_k = 1$ for all $k$.
}
\label{fig:cartoon}
\end{figure*}
 
Despite having different applications and underlying assumptions, an interesting observation is that the above randomised algorithms can be understood within the framework of randomised composite linear combination of unitaries (LCU). Specifically, they are all implemented by randomly sampling terms from the LCU formula~\cite{childs2012Hamiltonian,childs2018toward,berry2015Hamiltonian,berry2015simulating,chakraborty2024implementing} according to a specified probability distribution.
For example, although algorithms like PQS were not explicitly formulated using the LCU framework, the key idea is to decompose the interaction terms across different subsystems into local tensor-product types of operators, which is essentially an LCU decomposition. From an implementation standpoint, these decompositions are typically realised using Hadamard-test circuits~\cite{kitaev1995quantum}, further unifying them under the randomised LCU paradigm. Here, we explore the role of LCU employed in diverse contexts, and how the composition and randomisation play a common operational role across these seemingly disparate methods.

A consequence of randomisation is that, while it avoids the need for coherent LCU implementations, which reduces the qubit and gate counts, it sacrifices the ability to reconstruct the full quantum state. {As such, existing randomised algorithms are primarily constrained to estimating the expectation value of single Pauli operators only} (and to a greater extent, observables that admit an efficient Pauli decomposition). The subtleties are, in fact, twofold: (i) the state is not deterministically prepared, and (ii) the state appearing in the randomised LCU algorithms could be unphysical, e.g., in the form of $U \rho V ^{\dagger}$. {This naturally leads to the following question: {Can we go beyond estimating individual Pauli observables, while retaining the low-resource benefits of randomisation?}}

In this work, we show  how to estimate many observables simultaneously, as if we had actually prepared the (unphysical) state. To that end, we present a detailed construction of unbiased estimators based on the framework of random-sampling composite LCU. An observation is that they are built upon the Hadamard-test type circuits, which generate the effective state in the form of $\Phi(\rho) = U \rho V^{\dagger}$, as originally formulated in PQS~\cite{sun2021perturbative}. Here, we generalise it to the state appearing in the LCU formula $\prod_k U_k \rho \prod_k V_k^{\dagger}$, which is commonly used in Hamiltonian simulation and eigenstate property estimation tasks. \sun{Moreover,  we demonstrate how to effectively realise the state prepared by applying an operator that admits a composite LCU form.}
We explicitly show how to construct estimators that allow access to these effective states, enabling efficient observable estimation. From this perspective, it becomes evident that classical shadows  and randomised LCU are naturally compatible. The circuits for realising the effective state are presented in \autoref{fig:cartoon}, among which \autoref{fig:cartoon}(d) could be preferable in the fault-tolerant regime with frequent ancilla reset and measurement. 
Finally, as concrete applications, we present the construction of estimators for three representative use cases of randomised LCU in quantum simulation: High-precision Hamiltonian simulation by compensating for Trotter errors~\cite{zeng2022simple,wan2021randomized,yang2021accelerated}, eigenstate property estimation~\cite{sun2024high, zeng2021universal}, and large-scale real-time evolution simulated on smaller devices~\cite{sun2021perturbative}.
These examples correspond to the motivating scenarios for the design and use of LCU in quantum simulation discussed earlier. We present the gate complexity in tasks of dynamical and eigenstate property estimation.

 % most discuss how to extract the expectation value of Pauli operator
% generally aim to estimate a single observable at a time, rather than extract rich structural information about the quantum state itself.

% can consider estimating observable expectation values di-
% rectly, rather than preparing the full state via LCU.

% Recently, there has been significant progress in developing quantum algorithms designed for fault-tolerant quantum computing (FTQC) devices. 
% Among these, the random-sampling Linear Combination of Unitaries (LCU) framework is a very promising candidate for applications in the near-term and fault-tolerant quantum computing (FTQC) era.
% used technique in quantum algorithms, where the unitary is expressed as a weighted sum of simpler unitaries. 

% \section{Results}

% This work aims to provide a useful guide to two classes of applications which use Hadamard-test circuits as a building block.
% These include real-time evolution based on Trotterisaiton and eigenstate property estimation, where the effective eigenstate is realised via a random sampling method.

% \autoref{fig:cartoon}(b) shows a Hadamard-test-type circuit.

\section{Framework}

\subsection{Composite randomised LCU in quantum simulation}

The LCU framework~\cite{childs2012Hamiltonian,chakraborty2024implementing} is broadly used in quantum algorithms, such as ground state preparation \cite{Ge19} and real-time evolution~\cite{childs2018toward}.
The idea is to decompose the target unitary $U$ as a linear combination of unitaries
\begin{equation}
\label{eq:RTE_LCU}
    U = \sum_i \alpha_i U_i = \mu \sum_i \Pr(i) U_i
\end{equation}
where  $\mu =\sum_i \alpha_i$ is the $l_1$-norm of the coefficients and $\Pr(i)$ is the probability distribution over different $U_i$. Here, $ \alpha_i > 0$ is assumed and the phase is absorbed into the unitary $U_i$,  which could be some easy-to-implement unitaries such as Pauli operators $U_i \in (\pm 1) \times \{I, X,Y,Z\}^{\otimes n}$.

One way to implement the LCU method is by querying the Select and Prepare operations~\cite{childs2018toward}, which often incurs significant overhead due to the extensive use of controlled gates and additional ancilla qubits. As a more practical and efficient alternative, there is growing interest in realising \autoref{eq:RTE_LCU} through random sampling~\cite{faehrmann2022randomizing,zeng2021universal,sun2024high}. Looking at \autoref{eq:RTE_LCU}, the coefficient in the LCU formula $\Pr(i)$ can be interpreted as the probability for sampling the corresponding $U_i$, which, on average, results in an unbiased implementation of LCU $\mathbb{E}_i U_i = U$. The benefit of implementing through random sampling is that, if the target is the estimation of observable expectation values, we do not have to prepare the full state via the coherent implementation LCU. Instead, we can use a Hadamard-test-type circuit~\cite{kitaev1995quantum} to realise the state effectively, as used in Refs.~\cite{lin2022heisenberg,zeng2021universal,wang2023quantum,zhang2022computing}. The observable expectation is estimated by classically post-processing the measurement outcomes.

The sampling cost due to randomisation is related to the normalisation factor, which amplifies the variance by a factor of $\mu^2$, in a similar spirit to the cost in quantum error mitigation~\cite{Sugurureview,sun2021mitigating}. A direct decomposition of  $U$ may result in an exploding normalisation factor $\mu$. To keep the normalisation factor bounded, a strategy is to decompose a unitary operator that is close to the identity. In doing so, the normalisation factor, and thus the sampling cost, can be effectively controlled. \sun{Trotteration for real-time evolution, for example, manifests this idea, in which the total time $t$ is divided into $\nu$ segments. The techniques for reducing the cost incurred by randomisation are introduced in advanced dynamics simulation algorithms in  \cite{sun2021perturbative,wan2021randomized,yang2021accelerated,zeng2022simple}.}

% Therefore, in the context of quantum simulation applications,  

In order to apply randomised LCU for quantum simulation applications, it is often necessary to use a composite LCU formula as formulated and discussed in \cite{faehrmann2022randomizing,sun2024high}. 
As a concrete example, a product of the individual LCU formula described in \autoref{eq:RTE_LCU} takes the form of
\begin{equation}
\label{eq:product_LCU}
     U^{\nu} = \mu_T \sum_{\mathbf{i}} \Pr(\mathbf{i}) U_{\mathbf{i}}
\end{equation}
where $\mu_T = \mu^\nu$, each of the circuit instance $U_{\mathbf{i}} = \prod_{k=1}^\nu  U_{i_k}$ is labelled by $\mathbf{i} := \{i_1, ..i_\nu \}$  and the probability distribution associated with  $U_{\mathbf{i}}$ is $\Pr(\mathbf{i}) = \prod_{i_k} \Pr(i_k)$. By dividing the overall unitary into $\nu$ segments, the normalisation factor $ \mu_T$ can typically be bounded. For instance, following the method in \cite{zeng2022simple}, one can choose an appropriate $\nu$ such that the normalisation factor satisfies $\mu_T \leq 2$.
Therefore, \autoref{eq:product_LCU}  can also be understood as an LCU formula. 
The random sampling implementation of \autoref{eq:product_LCU} is very similar to that of \autoref{eq:RTE_LCU}, where the instance $\mathbf{i} $ is drawn from the corresponding probability distribution.

% The LCU form can be concatenated; for example, in simulating real-time evolution, we typically implement the Trotter formula $U_S$. The product $(U \dot U_S)^{\nu}$ can be similarly decomposed into an LCU form. We will discuss the structure of this concatenation in the context of specific applications.

% Without loss of generality
% For ease of notation, we denote the each of the circuit instance as 

% Look at the observable estimation for one instance $\mathbf{i} = \{ k \}_{k=1}^{\nu} $ in \autoref{eq:product_LCU} drawn from the probability distribution $\Pr(\mathbf{i})$, which is 
% \begin{equation}
% \label{eq:observ_LCU_product}
%     \tr (O U^{\nu} \rho (U^{\nu})^{\dagger} ) = \mu^2 \tr ( \prod_{k=1}^{\nu} U_k  \rho  \prod_{k=1}^{\nu}   V_k^{\dagger} O ).
%     \end{equation}
% To estimate a single observable $O$, this can be estimated via the  Hadamard-test circuit by introducing one ancillary qubit and measuring it on the ancilla, with the circuit shown in \autoref{fig:cartoon}(a) as discussed shortly after.

\subsection{ Quantum circuit realisation for the unphysical state}

% Generlaly speaking, this type of circuit can be realised via the circuit shown in \autoref{fig:cartoon}(d).

In the above section, the intermediate state generated in each sampled circuit instance takes the form of 
$\prod_{k=1}^{\nu} U_k    \rho  \prod_{k=1}^{\nu}  V_k^{\dagger}$. Here $U_k$ and $V_k$ represent the unitaries that are sampled in the $k$th segment, drawn independently from the same probability distribution by \autoref{eq:RTE_LCU}.
In this section, we first introduce the quantum circuit for generating an unbiased estimator for the above unphysical state. 
% Then, we discuss how to effectively realise the composite LCU via a quantum instrument.
% To start with, let us consider the realisation for the unphysical state.

Consider the quantum circuit illustrated in \autoref{fig:cartoon}(a), where we initialise the ancilla in $\ket{+}$, apply the controlled-unitaries followed by a phase gate ($Q^{b} = (S^{\dagger})^{b}$), and perform the measurement on the $X$ basis with measurement outcome $a = \{0, 1\}$.
Given the circuit setting $b$, the process of applying the unitary and measurements $\mathcal{E}_{a, b} = K_{a | b}  (\cdot)  K_{a | b}^{\dagger} $ is characterised by the Kraus operator, which maps an input state 
$\sigma$ to an output state of a measurement conditioned on a classical measurement outcome $a$.
The Kraus operator $ K_{a | b}$ takes the form of
\begin{equation}
    K_{a | b} = \frac{1}{2} \left((-i)^{b} (-1)^{a} \prod_{k=1}^{\nu} U_{k} + \prod_{k=1}^{\nu} V_{k} \right),
\end{equation} 
as derived in the Appendix.

\begin{proposition}
Denote the output state of the circuit in \autoref{fig:cartoon}(a) by $\sigma_{a,b}$ which is given by
$ 
    \sigma_{a,b} :=  \mathcal{E}_{a, b} (\rho) / \tr(  \mathcal{E}_{a, b} (\rho)).
$ 
The estimator  
\begin{equation}
\label{eq:estimator_v_rho_ab}
   \hat{v}_{a,b}  := 2 i^{b} (-1)^{a}  \sigma_{a,b},
\end{equation} is   unbiased, i.e., $\mathbb{E}_{a, b} \hat{v}_{a,b}  =  \prod_{k=1}^{\nu} U_k \rho \prod_{k=1}^{\nu} V_k^{\dagger} $.
\end{proposition}

In the Appendix, we provide the explicit form of the output state, from which one can verify the unbiasedness, $\mathbb{E}_{a, b} v_{a,b} =\sum_{a,b} \Pr(a|b)  v_{a,b}$ by computing the corresponding probability $\Pr(a|b)$. 

If we simply discard the measurement outcome by resetting the ancilla, this process reduces to a quantum channel. However, unlike the standard approach that traces out the ancilla, our method leverages the information from the ancilla measurement outcome, and thus enables us to effectively realise the unphysical state through quantum measurement combined with classical post-processing. It is important to note that \autoref{eq:estimator_v_rho_ab} does not imply that complete information of the state $\sigma_{a,b}$ is needed in order to obtain the unbiased estimator.
% we need to have access to the state $\rho_{a,b}$ or add additional operations. 
Rather, we only need a query to the quantum state $\sigma_{a,b}$. 
% without requiring detailed knowledge of $\sigma_{a,b}$. 
As we shall see in the next section, an unbiased estimator can be constructed from classical snapshots $\hat{\sigma}_{a,b}$ of the state through randomised measurements.

Next, we move to the case where a common unitary $U_S$ is applied deterministically within the segment, which corresponds to the scenario of simulating real-time evolution, where Trotterisation is typically employed. The circuit implementation is shown in \autoref{fig:cartoon}(c). An unbiased estimator can be similarly constructed as in \autoref{eq:estimator_v_rho_ab}:
\begin{equation}
\mathbb{E}_{{a}, {b}} \hat{v}_{{a}, {b}} = \prod_k U_k U_S \rho \prod_k U_S^\dagger V_k^\dagger. 
\end{equation}
To realise the time-evolution operator, $U_S$ can be the Trotter operator while  $V_k$ and $U_k$ are the sampled Pauli operators in the LCU decomposition of the Trotter-error compensation terms as proposed in \cite{zeng2022simple,wan2021randomized,yang2021accelerated}. The reader can see that there is no control over the $U_S$ operator in \autoref{fig:cartoon}(c).  The cost for implementing $U_S$ is usually larger than the randomly sampled unitary terms (see  \cite{sun2024high}), making it less costly. 

% In \autoref{fig:cartoon}(d), we actually consider a more general case where the unitaries $U_k$ and $V_k$ share a common unitary $U_S$, i.e., $U_k = \tilde{U}_k U_S$ and $V_k = \tilde{V}_k U_S$. This scenario usually arises in the simulation of real-time evolution, where Trotterisation is employed. For notational simplicity, we first present the formalism with $U_S$ being the identity. The corresponding circuit implementation will be discussed in \autoref{fig:cartoon}(c) and in the application section.

\subsection{ Quantum instrument realisation for composite LCU}

Next, we consider the scenario in which the objective is to implement the composite LCU operator $U^{\nu}$ acting on a quantum state. One straightforward approach to realise the composite LCU specified in \autoref{eq:product_LCU} is to use the circuit shown in \autoref{fig:cartoon}(a), perform measurements on the output of each circuit instance, and obtain the final result by averaging over the measurement outcomes.

However, a key distinction from the earlier discussion is that although the intermediate state generated in each individual instance is unphysical, the ensemble average over all instances $\mathbf{i}$ yields a physical state. This observation enables circuit variants in which the control operation is applied only within a much shorter period. Furthermore, from a resource-efficiency standpoint, realising this averaged state requires only a single circuit configuration—specifically, a single value of $b$—which further reduces implementation overhead. Many quantum algorithms, including those for eigenstate and dynamical property estimation, can be simplified by directly implementing the composite LCU, lowering the resource cost compared to naively implementing the unphysical terms individually.

Specifically, let us consider the circuit in \autoref{fig:cartoon}(d). In each of the segments in \autoref{fig:cartoon}(d), we add the ancilla initialised in $\ket{+}$, apply the controlled-unitaries ($U_{k}$ and $V_k$ randomly sampled from the distribution by \autoref{eq:RTE_LCU}) followed by a phase gate ($Q^{b_k} = (S^{\dagger})^{b_k}$), and perform the measurement on the $X$ basis with measurement outcome $a_k = \{0, 1\}$.
Here, $b_k = \{0, 1\}$ is generated uniformly at random which determines whether the phase gate is applied non-trivially. 
% The unitary $U_{k}$ and $V_k$ for $k = 1,..., \nu$ are randomly sampled from the distribution by \autoref{eq:product_LCU}.
The measurement outcome on the ancilla and the circuit setting is recorded as $(a_k, b_k )$.
Note that in \autoref{fig:cartoon}(d), we actually consider a more general case where the unitaries $U_k$ and $V_k$ share a common unitary $U_S$, i.e., $U_k = \tilde{U}_k U_S$ and $V_k = \tilde{V}_k U_S$.  For notational simplicity, we will present the formalism with $U_S$ being the identity. 
% The corresponding circuit implementation will be discussed in \autoref{fig:cartoon}(c) and in the application section.

The key feature of the circuit in \autoref{fig:cartoon}(d) is that the ancillary qubit is measured and reset within each segment. This measure-and-reset circuit is particularly useful in the FTQC regime, where frequently resetting the qubit is much more preferable than preserving it until the end of the computation.
% The circuits in  \autoref{fig:cartoon}(b) are special cases of \autoref{fig:cartoon}(d).
For example, when there is only one segment with $\nu = 1$, then it reduces to the Hadamard test circuit \autoref{fig:cartoon}(b).
For NISQ applications, we usually do not want to do reset operations. In this case, it is preferable to keep the ancilla until the end without any mid-circuit measurement as illustrated in \autoref{fig:cartoon}(c).

Let us formulate the process within each segment in a quantum instrument framework. 
Given the circuit setting $b_k$, the process of applying the unitary and measurements in the $k$th segment $\mathcal{E}_{a_{k}, b_{k}} = K_{a_{k} |b_{k}}  (\cdot)  K_{a_{k} |b_{k}}^{\dagger} $ is characterised by the Kraus operator, which maps an input state 
$\sigma$ to an output state of a measurement conditioned on a classical measurement outcome $a_k$.
The Kraus operator $ K_{a_{k} |b_{k}}$ is shown in the Appendix to take the form of
$ 
    K_{a_{k} |b_{k}} = \frac{1}{2} \left( (-i)^{b_{k}} (-1)^{a_{k}} U_{k} + V_{k} \right). 
$

The output state in each segment can be obtained iteratively. Let us denote the input normalised state before the $k$th segment as $\sigma^{(k-1)}$, the resulting state in the $k$th segment is given by
 $\sigma^{(k)} =   \mathcal{E}_{a_{k}, b_{k}} ( \sigma^{(k-1)}  ) /\tr( \mathcal{E}_{a_{k}, b_{k}} ( \sigma^{(k-1)}  ) )  $.
 Here, $ \sigma^{(k)} = \sigma^{(k)}_{ \{a_i\}_{i=1}^k, \{b_i\}_{i=1}^k }$ is an abbreviated notation for the output state as it represents an ensemble over configurations determined by the historical outcomes ${ (a_i, b_i) }_{k}$.

% Recall that in the previous section, we mainly focus on the intermediate state. In the following, we will establish the set-up and show how to directly construct the estimator for the state that is generated by the composite LCU form. 

Denote the final state generated from the circuit with $\{ b_k =0  \}$ in \autoref{fig:cartoon}(d)  as $ \sigma^{(\nu)}$, and record measurement outcomes $\{ a_k \}$. We have the following result.

\begin{proposition}

Suppose that $U^{\mu}$ is given by a composite LCU form defined in \autoref{eq:product_LCU} with a bounded normalisation factor $\mu_T $. Denote the sampled unitaries over the $k$ segments as $U_{\mathbf{i}}$ and $U_{\mathbf{j}}$, which are independently and randomly drawn according to the probability distribution defined in \autoref{eq:product_LCU}.
Define the estimator as
$
   \hat{v}^{(\nu)}_{\mathbf{a}} :=  (-1)^{\sum_{k=1}^{\nu} a_k}  \sigma^{(\nu)},$  with   $\mathbf{a} := \{ a_k\}_{k=1}^{\nu}$.
Then we have   $ \mu_T^2 \mathbb{E}_{\mathbf{i},\mathbf{j}} \mathbb{E}_{\mathbf{a}} \hat{v}^{(\nu)}_{\mathbf{a}} = U^{\nu} \rho (U ^{\nu})   ^{\dagger}  $.
    
\end{proposition}

Here, the normalisation factor $\mu_T$ is solely determined by the composite LCU formula, which usually can be controlled by setting appropriate $\nu$ such that $\mu_T = \mathcal{O}(1)$.
We note that the composition by repetition of $U^{\mu}$ is merely one example. More generally, different LCU formulas can be applied to each segment; for instance, using $\prod_{i=1}^{\nu} U(t_i)$ with varying $t_i$ across segments.
 
This result demonstrates how to directly construct the estimator for the state generated by the composite LCU form, in contrast to that in the previous section which focused on the intermediate state.
A difference is that the circuit is fixed, i.e. no randomisation over $b_k$ to realise the composite LCU. This simplification is made possible by exploiting the Hermiticity of the target state (which can be understood as time-reversal symmetry).  The proof for realising the composite LCU based on the circuit in \autoref{fig:cartoon}(d) is technically involved and will be presented in the Appendix, whereas the result based on \autoref{fig:cartoon}(a) is more straightforward.

% We also note that the circuit in \autoref{fig:cartoon}(d) is particularly useful when frequent qubit resetting is preferable in the FTQC regime, whereas circuits in \autoref{fig:cartoon}(b) are more suited to applications in the NISQ era. 

% In all these cases, the unbiased estimator is constructed for the unphysical state that arises in the composite LCU formulation given by \autoref{eq:product_LCU}. This is a general result that unifies the previous estimations by variants of the Hadamard test circuit.

% Denote the unnormalised state in the  $k$th segment  as $\rho^{(k)}$. This state, indeed, depends on the outcome $a_{k}$ and $b_{k}$ on the ancilla. 

\subsection{Shadow estimation for many observables}

In the above discussion, we primarily focused on how to realise the non-completely-positive (non-CP) map acting on a state, given access to the state $\sigma_{a,b}$ or $\sigma^{(\nu)}$. We now explicitly discuss how to construct an unbiased estimator for the resulting composite state and how to enable observable estimation based on it.
It is very natural to introduce a shadow of the state that is obtained from different types of circuits in \autoref{fig:cartoon}. 
We apply the shadow estimation method proposed by Huang et al~\cite{huang2020predicting}. We apply a unitary operator $R$ chosen randomly from a tomographically complete unitary set $\mathcal{C}$, which can be a set of random Clifford or Pauli operators. Recent works have shown that the random unitaries can be implemented in low depth~\cite{schuster2024random,bertoni2024shallow}. The measurement outcome is obtained by measuring on the system qubits, which is denoted as $\vec{z} := \vec{z}^{(b)} \in Z_2^n $ with $b = 0 $ or $1$. For simplicity, the superscript $b$ is omitted when there is no ambiguity.

With the bitstring $\vec{z}$ obtained from randomised measurements, we can get an unbiased snapshot of the state $ \rho^{(\nu)}$ as
\begin{equation}
    \hat{\rho}_{a,b}(R, \vec{z})
=  \mathcal{M}^{-1} \left( R^\dagger |\vec{z}\rangle \langle \vec{z}| R \right)
\label{eq:shadow_estimator_v_a}
\end{equation} with $\mathbb{E}_{R, \vec{z}} \left( \hat{\rho}_{a,b} (R, \vec{z}) \right) = \sigma_{a,b}$,  where $\mathcal{M}^{-1}  $ is a invertible linear map determined by $\mathcal{C}$.
Define the estimator as
$ 
    \hat{v}_{a,b}(R, \vec{z}) =  i^{b} (-1)^{ a}   \hat{\rho}_{a,b}(R, \vec{z})
$, which is unbiased, $$  \mathbb{E}_{{a}, {b}, R, \vec{z}}  \hat{v}_{a,b}(R, \vec{z}) =  \prod_{k = 1}^{\nu} U_k \rho \prod_{k = 1}^{\nu}  V_{k}^{\dagger}. $$
Define $\hat{o}_m :=  \tr( O_m \hat{v}_{a,b}) $ for observable $O_m$.
It is easy to check that
$ 
   \mathbb{E}_{a,b, R, \vec{z}}\hat{o}_m =  \tr( O_m  \prod_{k = 1}^{\nu} U_k \rho \prod_{k = 1}^{\nu}  V_{k}^{\dagger} ) 
$.

% As such, 
% \begin{equation}
%      \tr ( \mathcal{M}^{-1}(O)  \hat{\rho}_{a} ) =  \sum_b \tr (O  i^b (-1)^a \mathcal{M}^{-1} (\hat{v}_a^{(0)}(R, \vec{z}))).
% \end{equation}
% just  need to compute 
% \begin{equation}
%     \tr ( \mathcal{M}^{-1}(O)   R^\dagger |\vec{z}\rangle \langle \vec{z}| R  )
% \end{equation}
% due to self-duality of $\mathcal{M}$.

The required sampling number is related to the variance of $\hat{o}_m $, which is bounded by  the shadow norm~\cite{huang2020predicting} of the observable 
\begin{equation}
    \Var(\hat{o}_m) \leq 2 \| O \|_{\rm shadow}^2  - \tr ( O_m \prod_{k = 1}^{\nu} U_k \rho \prod_{k = 1}^{\nu}  V_{k}^{\dagger})^2.
\end{equation}
Here, the shadow norm~\cite{huang2020predicting} of the observable takes the form as
\begin{equation}
\|O\|_{\text{shadow}}^2 = \max_{\sigma } \mathbb{E}_{P  } \sum_{\vec{z} \in Z_2^n} 
\langle \vec{z} | P \sigma P^\dagger | \vec{z} \rangle 
\langle \vec{z} | P \mathcal{M}^{-1}  (O) P^\dagger | \vec{z} \rangle^2.
\end{equation}
where $P = U^\nu$ is a unitary.
The derivation of the variances of $\hat{o}_m$ uses the self-duality of $\mathcal{M}$:
$ 
    \tr (\mathcal{M}^{-1}(O) \rho ) =   \tr (\mathcal{M}^{-1}( \rho)  O),
$
where $\rho$ is a state $\rho \in \mathcal{H}(2^n)$. The proof is presented in the Appendix.
We have picked up the circuit in \autoref{fig:cartoon}(a) for illustration. The workflow for other circuits is rather similar.

% A compromised way to get $\tr( O U \rho V ^{\dagger}) $ is to use the circuit in \autoref{fig:cartoon}(b). The measurement circuit requires an additional state preparation circuit $U_p$.

\section{Shadow Estimation in Randomised LCU-Based Quantum Simulation } 

The above section has established the general framework for the composite randomised LCU and constructed the unbiased estimator. In this section, we discuss the role of randomised LCU in two representative applications in quantum simulation: dynamical and eigenstate property estimation. We shall see that starting from a simple initial state, randomised LCU is used to effectively (not deterministically) generate the time-evolved state and the eigenstate. We demonstrate how these applications can be realised under the same framework, highlight the role of LCU in various applications, and present the corresponding depth and sample complexity.

\subsection{Real-time evolution}

To simulate real-time evolution, many quantum algorithms rely on Trotterisation. Among them, the Trotter-LCU and PQS approaches are two representative methods. Trotter-LCU aims to achieve high-precision real-time evolution, while PQS targets simulating larger quantum systems using smaller quantum devices with fewer qubits. Although designed for different purposes, these two methods share similarities in both the implementation perspective and the role of the LCU, which are fundamentally aligned. Both can be understood within the framework of randomised composite LCU.

In the task of dynamical property estimation, the initial state is evolved under $\prod_{\delta t}^T \mathcal{V}(\delta t) \circ \mathcal{U}(\delta t)$ where the total time $T$ is divided into $\nu = T/\delta t$ segments. In PQS, we decompose the interaction non-local term. In Trotter-LCU, the essential idea is to decompose the Trotter-error term, which can also be regarded as an 'interaction' term,  into Pauli operators. 
% In both cases, both Trotterisation and the interaction term are effectively realised for the purpose of compensating for error and enabling larger simulations with local operations only.

We now explicitly demonstrate how large-scale simulation by PQS fits into this framework. In each step, the interacting channel $ \mathcal{V} = e^{-i H_{\rm int} \delta t} (\cdot) e^{i H_{\rm int} \delta t}$ is governed by the interacting Hamiltonians $H_{\rm int}  \in \mathcal{H}( 2^{n \times L }))$ acting on $n \times L$ qubits.  Suppose  $H_{\rm int} = \sum_i p_i P_i $ with $P_i$ being Pauli operators.
The interacting channel, as per PQS, is decomposed into local generalised quantum operations (i.e. $\Phi(\rho) = U \rho V^{\dagger}$) as
\begin{equation}
\begin{aligned}
    \mathcal{V}(\delta t) (\rho) &= \mathcal{I} -   i  (  H_{\rm int} \rho -  \rho H_{\rm int} ) \delta t + \mathcal{O}(\delta t^2).\nonumber
\end{aligned}
\end{equation}
 By factoring out a normalisation constant, it is essentially an LCU form
\begin{equation}
     \mathcal{V}(\delta t) (\rho) = \mu \left( \alpha_0 {\mathcal{I}} +  \sum_i \left(  \tilde{P}_i \rho +  \rho \tilde{P}_i^{\dagger} \right) \alpha_i \right) + \mathcal{O}(\delta t^2) 
\end{equation}
where $\mu$ is the normalisation factor, the phase $i$ is incorporated into $P_i$, $ \tilde{P}_i := (-i)P_i$,  i.e., $ \alpha_i := p_i \delta t / \mu$, and $\alpha_0 = 1 - \sum_i |\alpha_i|$.

Compared to unitary decomposition, the PQS approach involves a decomposition of the quantum channel rather than a unitary operator. At first glance, this may suggest a loss of symmetry. However, an observation is that the paired state takes the form of $\tilde{P}_i \rho +  \rho \tilde{P}_i^{\dagger}$ with $\tilde{P}$ being either Pauli or identity operator which can be realised with a single circuit configuration by fixing $b_k = 0$. Notably, it can be shown that implementations with $b_k = 0$ and $b_k = 1$ are equivalent. The difference lies in the treatment of the phase: the circuit setting with $b_k = 1$ corresponds to an implementation that does not incorporate the phase into~$\tilde{P}_i$. 
% Therefore, \autoref{prop:Trotter_LCU_estimator} naturally extends to larger-scale simulation by PQS.

PQS can be regarded as a more general application that extends to many copies.
When only one copy is considered, it reduces to the Trotter-LCU. 
Specifically, this correspondence becomes evident when interpreting the interaction term $\mathcal{V}(\delta t)$ as the Trotter error term, which can be explicitly computed as $\mathcal{V}(\delta t) = \mathcal{U}(\delta t)  \tilde{\mathcal{S}}(\delta t) $  where $\tilde{\mathcal{S}} = U_S^{\dagger} (\cdot) U_S$.

Previous works have mainly discussed how to estimate the expectation value of a single observable by measuring it on the ancilla. Having established the estimate for the unphysical state, we are able to estimate the expectation values of many observables by using the information on the system qubits.

% Below, we elaborate on the implementation of Trotter-LCU and PQS.
% For a unitary $U$ expressed in an LCU form, suppose that $U_{k_1}$ and $U_{k_2}$ are generated randomly and independently based on the probability defined in \autoref{eq:RTE_LCU}. It is straightforward to have  $ \mathbb{E}_{k_1, k_2}   U_{k_1} \rho U_{k_2}^{\dagger} =   U \rho U^{\dagger}  $. 
% Provided its LCU formula, we can construct an unbiased estimator for $U \rho U^{\dagger}$, with the shadow estimation  based on \autoref{eq:shadow_estimator_v_a}

 Suppose the target unitary is decomposed into $\nu $ segments, $U(T) = ( U(\delta t))^\nu$, and has a composite LCU form with $\mu_T = \mu^{\nu}$ described in \autoref{eq:product_LCU}.
Denote the circuit instances as $U_{\mathbf{i}}$ and $U_{\mathbf{j}}$ labelled by $\mathbf{i} $ and $\mathbf{j} $,  which are drawn uniformly and independently from the corresponding probability distribution.
We consider two cases. Case I: Measurement via circuit in \autoref{fig:cartoon}(d), and Case II: Measurement via circuit in \autoref{fig:cartoon}(a) or (c). 

\begin{proposition}[Unbiased estimator for dynamic property estimation]
\label{prop:Trotter_LCU_estimator}

% Suppose that the circuit is generated by $U = U_{k_1}$ and $V = U_{k_2}$ where $U_{k_1}$ and $U_{k_2}$ are generated randomly and independently based on the probability defined in \autoref{eq:RTE_LCU} with the initial state $\rho$. 
Case I.  Circuit with frequent means of reset in \autoref{fig:cartoon}(d). 
Set the quantum instrument with $\mathbf{b}=0$.
After randomised measurements by applying unitaries $ R$ chosen from $\mathcal{C}$, denote the snapshot of the final output state as $\rho^{(\nu)}(R, \vec{z}) $ defined in \autoref{eq:shadow_estimator_v_a}.
The estimator, defined as
\begin{equation}
   \hat{v}^{(\nu)}_{\mathbf{a}, \mathbf{b}=0}(R, \vec{z})  := (-1)^{\sum_{k=1}^{\nu} a_k}  \rho^{(\nu)}(R, \vec{z}), 
\end{equation} can be used to unbiasedly generate the real-time-evolved state, i.e., 
\begin{equation}
 \mu_T^2 \mathbb{E}_{ \mathbf{i}, \mathbf{j}} \mathbb{E}_{R, a, \vec{z} }    \hat{v}^{(\nu)}_{\mathbf{a}, \mathbf{b}=0}(R, \vec{z})= U(T) \rho U(T)^{\dagger}.\nonumber
\end{equation}

\end{proposition}
 
This result holds true for both the Trotter-LCU and PQS scenarios.
The measurement protocol with an always-on ancilla is nearly identical. For the reader’s convenience, we briefly summarise the procedure for Case II below.
Set the quantum circuit in   \autoref{fig:cartoon}(a,c).
After randomised measurements by applying unitaries $ R$ chosen from $\mathcal{C}$, denote the snapshot of the final output state as
\begin{equation}
  \hat{\rho}_{a, b}(R, \vec{z})
=  \mathcal{M}^{-1} \left( R^\dagger |\vec{z}\rangle \langle \vec{z}| R \right).\nonumber
\end{equation} which  is unbiased
$ 
\mathbb{E}_{R, \vec{z}} \left( \hat{\rho}_{a, b} \right) = \rho_{a, b}
$. For the case of real-time evolution, the estimator is defined as
\begin{equation}
    \hat{v}_{a,b=0}(R, \vec{z}) =   (-1)^a \hat{\rho}_{a,b=0}(R, \vec{z}).\nonumber
\end{equation} is an unbiased estimator for the real-time-evolved state, i.e. 
$
 \mu_T^2 \mathbb{E}_{ \mathbf{i}, \mathbf{j}} \mathbb{E}_{R, a, \vec{z} }    \hat{v}_{{a}, {b}=0}(R, \vec{z}) = U(T) \rho U(T)^{\dagger}.\nonumber
 $

% So far, we have discussed how to realise the Trotter error term. 
% Then, putting it together by considering the shared Trotter formula $S$, we can use \autoref{fig:cartoon}(c) to realise an unbiased estimator for the time-evolved state.

For simulating real-time evolution, the circuit depth by Trotter methods scales polynomially in the target precision, more specifically, $\mathcal{O} ( t^{1 + 1/2k} \varepsilon^{-1/2k} )$ when we use the $k$th order Trotter formula. By incorporating Trotter-LCU with shadow estimation, we no longer need to prepare the state; instead, we can simply estimate the expectation values of many observables. In this case, the maximum circuit depth is logarithmic in precision $\mathcal{O} ( t^{1 + 1/4k} \log(\varepsilon^{-1}) ) $, showing advantages over Trotterisation-based methods.

In the above discussion, we primarily focused on approximating the unitary with zero error. It is straightforward to extend to cases where there is a truncation error in the approximation by a truncation of the LCU formula, i.e. $(\mu, \varepsilon)$-LCU formula, $\tilde{U} = \mu \sum_k \Pr(k) U_k$ which satisfies $\| U -  \tilde{U} \| \leq \varepsilon$  as introduced in \cite{zeng2022simple}. The results can be derived in the same way.

\subsection{Eigenstate property estimation}

% Starting from an initial state with a nonzero overlap with the eigenstate, the question is how to get an  estimation of observable expectation values on the eigenstate.

The other application is about estimating the properties of the eigenstate $| E_j \rangle$ of a Hamiltonian, provided an initial state with a nonzero overlap with the eigenstate~\cite{lin2020near,dong2022ground}.
% If the eigenstate  $| E_j \rangle$ can be prepared deterministically, then by applying the standard shadow, we can estimate the property.
However, $|E_j\rangle$ is usually hard to prepare and thus not accessible directly. 
% The question is how to get the shadow of $|\phi_j\rangle$ without direct access to $|\phi_j\rangle$? 
As discussed in \cite{zeng2021universal}, the eigenstate can be effectively realised by evolving under invariant time evolution. The key is to realise this non-unitary evolution $g_\tau(H - \omega) := e^{-\tau^2 (H^2 - \omega)^2} $. Let us consider the unnormalised state $ |\phi_j\rangle := g_\tau(H - \omega) \ket{\psi_0}  $ which is referred to as the approximate eigenstate, which becomes the exact eigenstate in the long time limit $| E_j \rangle = \lim_{ \tau \rightarrow \infty} g_\tau(H - \omega) \ket{\psi_0} $, when we take $\omega = E_j$. The time complexity required is discussed in \cite{sun2024high}.

% Below, we apply the spectral filter method to prepare the effective state $|E_j\rangle$ based on randomised LCU  and 
Below, we will discuss how to estimate eigenstate properties by incorporating shadow estimation into randomised LCU.
With the LCU decomposition, the unnormalised approximate eigenstate  
\[
|\phi_j\rangle 
= \int dx\, \Pr(x)\,  e^{-i\tau x \omega} e^{i\tau x H} |\psi_0\rangle
\] and the observable expectation on the unnormalised state 
\begin{equation}
\begin{aligned}
&\langle \phi_j | O  | \phi_j \rangle
= \langle \psi_0 | g_\tau(H - \omega) O\,  g_\tau(H - \omega)| \psi_0 \rangle \\
&=  \int d \mathbf{x} \Pr(\mathbf{x}) e^{-i\tau(x_1 - x_2)\omega}
\, \mathrm{Tr}\left(O\, U(x_2)\, \rho\, U^\dagger(x_1)\right)
\end{aligned}
\end{equation}
where    $\mathbf{x}:= (x_1,x_2)$, $\Pr(\mathbf{x}):= \Pr(x_1) \Pr(x_2)$ and $U (x_i) = e^{-i x_i H}$.
Since $x_1$ and $x_2$ are symmetrically interchangeable, the observable expectation value can be symmetrised by absorbing the phase into the unitary. This leads to the modified unitary $\tilde{U}(x) := e^{i x_i \tau \omega} U(x)$. \sun{The advantage is that the circuit instance is fixed, and thus the sampling number is reduced by a factor of $4$ compared to that in existing works.}

The expectation value of the observable on the approximate eigenstate is given by
$ 
    \braket{O} = {\tr( O |\phi_j\rangle \langle \phi_j| )} / { \tr(|\phi_j\rangle \langle \phi_j|) }
$.
The key question is how to construct an estimator  $\hat{v}$ explicitly such that 
$ 
 \mathbb{E}    \hat{v} = |\phi_j\rangle \langle \phi_j|
$.

\begin{proposition}[Unbiased estimator for eigenstate property estimation]

Define the estimator as
\begin{equation}
\label{eq:estimator_RLCU}
    \hat{v}_{a,b}(R, \vec{z}) =  (-1)^a \hat{\rho}_{a, b=0}(R, \vec{z})
\end{equation}
where   $ \hat{\rho}_{a, b = 0}(R, \vec{z}) = \mathcal{M}^{-1} (R^\dagger |\vec{z}\rangle \langle \vec{z}| R)$, which is unbiased for the operator $v$, i.e., 
$ 
 \mathbb{E}_{\mathbf{x}}  \mathbb{E}_{ a, b, R, \vec{z}}   \hat{\rho}_{a, b}(R, \vec{z}) = |\phi_j\rangle \langle \phi_j| \nonumber
$
where  $ |\phi_j\rangle  = g_\tau(H - \omega) |\psi_0\rangle $.
    
\end{proposition}

This can be seen as follows.
\begin{equation}
\begin{aligned}
    \mathbb{E}_{\mathbf{x}}  \mathbb{E}_{ a, b, R, \vec{z}}   \hat{\rho}_{a, b}(R, \vec{z}) &=  \mathbb{E}_{\mathbf{x}} \tr(O \tilde{U}(x_1) \rho \tilde{U}(x_2)^{\dagger})  \\
    &= |\phi_j\rangle \langle \phi_j|.\nonumber
\end{aligned}
\end{equation}
With the unbiased estimator, we can estimate the expectation values of many observables on the eigenstate $|E_j \rangle$ approximated by $\ket{\phi_j}$
By incorporating the shadow estimation method into the random-sampling spectral filter method~\cite{zeng2021universal,sun2024high}, it is straightforward to arrive at the gate complexity for eigenstate property estimation, which is stated below.

\begin{corollary}[Eigenstate property estimation]
 To guarantee that the maximum estimation error of the expectation values of $M$ observables on the eigenstate within an error $\varepsilon$, with a failure probability $\delta$, Suppose we set the maximum time  $\mathcal{O} ( \Delta_j^{-1} \log(\varepsilon^{-1}))$  and the
number of measurements $\mathcal{O} ( \varepsilon^{-2} \| O\|_{\rm shadow} \log(M/ \delta) ) $, where $\Delta_j : = \min(E_{j+1} - E_j, E_{j} - E_{j-1})$ is the energy gap.

\end{corollary}

From now on, we can see that by using the unbiased estimator defined by \autoref{eq:estimator_RLCU},  the
sample complexity for estimating multiple observables' expectation values scale logarithmically in the number of observables $\mathcal{O} ( \log(M ) \| O\|_{\rm shadow}  ) $ while the   circuit depth  complexity scales logarithmically in $\varepsilon$,  $\mathcal{O} ( \Delta_j^{-1} \log(\varepsilon^{-1}))$.

% It may show advantages in situations where coherent implementation of the algorithms is required. 

\section{Discussion}

In this work, we explicitly demonstrate how composite LCU can be realised and how observables can be estimated simultaneously and efficiently. To achieve this, we introduce generalised Hadamard test circuits (\autoref{fig:cartoon}(a,c)) and their variants (\autoref{fig:cartoon}(d)), tailored to different use cases depending on whether the ancilla is kept active or frequently reset.
Using these circuits, we construct unbiased estimators for effective states of the form $U \rho V^\dagger$ and their generalisations across various tasks. This effectively realises a  non-CP map.
It is interesting to compare our approach with dissipative simulation. A key distinction lies in the mechanism: dissipative simulations implement Lindblad operators via completely positive and trace-preserving (CPTP) maps, whereas our method introduces an additional label on the output state, enabling the generation of non-Hermitian states in a controlled manner.
Beyond non-Hermiticity, the quantum instrument framework enables the realisation of a non-CP linear map in a random sampling way. An interesting future work is to design a dissipator that does not rely on CPTP maps, for example, a two-sided Lindblain can be used in the simulation of non-Markovian process~\cite{yang2023efficient}. It may also have direct relevance to applications in linear algebra.

% We have developed a unified framework for estimating observables from unphysical states generated in a randomised linear combination of unitaries algorithms. We explicitly construct unbiased estimators for effective states of the form $U \rho V^\dagger$ and their generalisations for different tasks. 

Based on the composite randomised LCU formula, we discuss how to estimate multiple observables efficiently. 
This composite approach bridges various quantum simulation methods, such as Hamiltonian simulation, spectral filter methods, and PQS, under a common operational framework, aligning with the spirit of early FTQC applications. As a by-product, the unification of PQS and Trotter-LCU can be naturally established within this framework. 
% For real-time evolution by either Trotter-LCU and PQS, we can pair the state and makes it Hermitian, such that only one circuit set and real-valued estimator is needed. 
In both Trotter-LCU and PQS approaches to real-time evolution, the intermediate state can be symmetrised to enforce Hermiticity (see Appendix). This enables the use of a single set of circuits and results in a real-valued estimator.
% \sun{ In contrast, for PQS or eigenstate-related tasks, a complex-valued estimator is necessary, as the imaginary part of the state plays a crucial role.}
As we demonstrate in the context of dynamical and eigenstate property estimation, our approach preserves the high-precision performance of existing algorithms, while enabling the simultaneous estimation of multiple observables. This is achieved by performing measurements on the system qubits, resources that are often discarded in previous Hadamard-test-based quantum algorithms. Given the unbiased estimator, advanced observable estimation methods, such as qubit-wise grouping~\cite{wu2021overlapped,yen2023deterministic} and derandomised classical shadows~\cite{huang2021efficient}, are compatible in this randomised LCU scenario and can thus be employed to further reduce the measurement cost. This is particularly important for quantum chemistry applications in the measurement cost scales quartically $\mathcal{O}(N^4)$ with the number of orbitals~\cite{patel2025quantum}. As shown in this work, the randomised LCU method combined with efficient measurement schemes offers favourable scaling in circuit depth and efficient observable estimation, making it well-suited for dynamical and eigenstate tasks in many-body physics.
In addition,  the framework is applicable to a broader class of distributed quantum computation (DQC)~\cite{parekh2021quantum}, in which PQS can be viewed as a specific instance. For example, observable estimation in hybrid tensor networks~\cite{yuan2020quantum,schuhmacher2025hybrid}, another form of DQC, can be incorporated into the measurement framework.

\vspace{6pt}
\emph{Note added---} We came across  a similar work posted on arXiv recently~\cite{faehrmann2025shadow} which considers the shadow of the Hadamard test circuit in \autoref{fig:cartoon}(b).

% \begin{acknowledgments}

% \end{acknowledgments}

% \newpage
\appendix

\widetext

% \sun{Below shows the other version  ---- more complicated and not used any more ---}

% \begin{equation}
%     \begin{aligned}
%         &\mathbb{E}_{a_{\nu}, b_{\nu}, ..., a_1, b_1} i^{\sum_{k=1}^{\nu} b_k} (-1)^{\sum_{k=1}^{\nu} a_k} v^{(\nu)}   \\
%         &=         \mathbb{E}_{a_{\nu-1}, b_{\nu-1}, ..., a_1, b_1} (-i)^{\sum_{k=1}^{\nu-1} b_k} (-1)^{\sum_{k=1}^{\nu-1} a_k} \mathbb{E}_{a_{\nu}, b_{\nu}} K_{a_{\nu} |b_{\nu}} v^{(\nu-1)} K_{a_{\nu} |b_{\nu}}^{\dagger} \\
%         &=         \mathbb{E}_{a_{\nu-1}, b_{\nu-1}, ..., a_1, b_1} (-i)^{\sum_{k=1}^{\nu-1} b_k} (-1)^{\sum_{k=1}^{\nu-1} a_k} \tr(v^{(\nu-1)}) \mathbb{E}_{a_{\nu}, b_{\nu}} K_{a_{\nu} |b_{\nu}} \rho^{(\nu-1)} K_{a_{\nu} |b_{\nu}}^{\dagger} \\
%         &=         \mathbb{E}_{a_{\nu-1}, b_{\nu-1}, ..., a_1, b_1} (-i)^{\sum_{k=1}^{\nu-1} b_k} (-1)^{\sum_{k=1}^{\nu-1} a_k} \tr(v^{(\nu-1)}) U_{\nu}  \rho^{(\nu-1)} V_{\nu}^{\dagger} \\
%         &=        U_{\nu}  \left(  \mathbb{E}_{a_{\nu-1}, b_{\nu-1}, ..., a_1, b_1} (-i)^{\sum_{k=1}^{\nu-1} b_k} (-1)^{\sum_{k=1}^{\nu-1} a_k} \tr(v^{(\nu-1)}) \rho^{(\nu-1)} \right) V_{\nu}^{\dagger}  \\
%          &=        U_{\nu}  \left(  \mathbb{E}_{a_{\nu-1}, b_{\nu-1}, ..., a_1, b_1} (-i)^{\sum_{k=1}^{\nu-1} b_k} (-1)^{\sum_{k=1}^{\nu-1} a_k} v^{(\nu-1)} \right) V_{\nu}^{\dagger}  \\
%          &=        U_{\nu} U_{\nu-1}... U_{1}  v^{(\nu-1)}  V_{1}^{\dagger}  V_{2}^{\dagger}... V_{\nu}^{\dagger} \\
%     \end{aligned}
% \end{equation}

\section{Output state generated by the Hadamard-test type of circuit }

Let us start with the Hadamard-test circuit depicted in \autoref{fig:cartoon}(b).
The unnormalised state generated by the Hadamard-test circuit with no phase gate (before measurement) is

\begin{equation}
    \begin{aligned}
    \frac{1}{4} \big(  |0\rangle\langle 0| \left( V \rho V^\dagger + U \rho V^\dagger + V \rho U^\dagger + U \rho U^\dagger \right)  
+  |0\rangle\langle 1| \left( V \rho V^\dagger + U \rho V^\dagger - V \rho U^\dagger - U \rho U^\dagger \right) \\
+  |1\rangle\langle 0| \left( V \rho V^\dagger - U \rho V^\dagger + V \rho U^\dagger - U \rho U^\dagger \right)  
+   |1\rangle\langle 1| \left( V \rho V^\dagger - U \rho V^\dagger - V \rho U^\dagger + U \rho U^\dagger \right) \big).
    \end{aligned}
\end{equation}

The unnormalised state generated by the Hadamard-test circuit with the inverted phase gate $S^{\dagger}$  (before measurement) is
\begin{equation}
    \begin{aligned}
        \frac{1}{4} ( 
|0\rangle\langle 0| \left( V \rho V^\dagger + \textcolor{red}{(-i)} U \rho V^\dagger + \textcolor{red}{i} V \rho U^\dagger + \textcolor{red}{i(-i)} U \rho U^\dagger \right) 
+ \; |0\rangle\langle 1| \left( V \rho V^\dagger + \textcolor{red}{(-i)} U \rho V^\dagger - \textcolor{red}{i} V \rho U^\dagger - \textcolor{red}{i(-i)} U \rho U^\dagger \right) \\
+ \; |1\rangle\langle 0| \left( V \rho V^\dagger - \textcolor{red}{(-i)} U \rho V^\dagger + \textcolor{red}{i} V \rho U^\dagger - \textcolor{red}{i(-i)} U \rho U^\dagger \right) 
+   |1\rangle\langle 1| \left( V \rho V^\dagger - \textcolor{red}{(-i)} U \rho V^\dagger - \textcolor{red}{i} V \rho U^\dagger + \textcolor{red}{i(-i)} U \rho U^\dagger \right) ). \\
    \end{aligned}
\end{equation}
Therefore, given the measurement result $a = 0$ or $1$ on $A$, the unnormalised state of the system $B$ can be expressed as 
\begin{equation}
    \rho_{a, b} = \frac{1}{4} ( (-i)^b  (-1)^a U + V ) \rho (  i^b (-1)^a U^{\dagger} + V^{\dagger} ).
\end{equation}

The probability of getting the measurement outcome $a$ given the circuit setting $b  = 0$ is 
\begin{equation}
    \Pr(a|b = 0) = \frac{1}{2} \left ( 1 + (-1)^a  \RE (\tr(U \rho V^{\dagger})) \right).
\end{equation}
Similarly, the probability of getting the measurement outcome $a$ given the circuit setting with applying the inverted phase gate (i.e. $b = 1$) is 
\begin{equation}
    \Pr(a|b = 1) = \frac{1}{2} \left ( 1 + (-1)^a  \IM (\tr(U \rho V^{\dagger})) \right).
\end{equation}

The normalised state of the system $B$ is thus given by 
\begin{equation}
    \sigma_{a,b} = \frac{\rho_{a, b}}{\tr(\rho_{a, b})} =    \frac{1}{4 \Pr(a|b )} ( (-i)^b  (-1)^a U + V ) \rho (  i^b (-1)^a U^{\dagger} + V^{\dagger} )
\end{equation}
with  $\Pr(a|b)  = \tr(\rho_{a,b})$.
We can see that by taking the average over $a$, for $b = 0$,  we have
\begin{equation}
  \mathbb{E }_a (-1)^a  \sigma_{a, b = 0} =  \frac{1}{2} (U  \rho V^{\dagger} + V  \rho U^{\dagger} )
\end{equation}
while for $b = 1$ we have 
\begin{equation}
  \mathbb{E}_{a} i^b (-1)^a \sigma_{a, b = 1} =  \frac{1}{2} (U  \rho V^{\dagger} - V  \rho U^{\dagger} )
\end{equation}

With this result, one can extend to the case with multiple unitaries. When the average is taken over $b$ (i.e. the circuit instance is randomly generated) for the estimator $\hat{v}_{a, b}$ defined in \autoref{eq:estimator_v_rho_ab}, we have 
\begin{equation}
  \mathbb{E }_{a, b} 2 i^b (-1)^a  \sigma_{a,b} = \sum_b 2 \Pr(b)  \mathbb{E }_{a} i^b  (-1)^a  \sigma_{a, b} =  \prod_{k=1}^\nu  U_k  \rho \prod_{k=1}^\nu V_k^{\dagger}
\end{equation}
where $\Pr(b) = \frac{1}{2}$. Proposition 1 is thus proven. Another way to define the estimator is
\begin{equation}
    \hat{v}_{a} = \sum_{b  \in \{0, 1\}} i^b (-1)^a  \sigma_{a,b},
\end{equation}
which is equivalent to the expression in \autoref{eq:estimator_v_rho_ab} in expectation.

The above simple case can be regarded as $\nu = 1$. 
We shall use this to prove the unbiasedness of general cases with multiple segments. 
However, the representation when we consider the composition becomes a bit complicated.
Before proceeding, let us examine the symmetry properties of the state that emerge when the average is taken over $a_k$ with fixed $b_k$.

\section{The symmetry in composite LCU and  the effective realisation with different circuits}

In our work, we consider the realisation of LCU by two types of circuits: (1) circuits where the ancilla is kept alive until the end, as shown in \autoref{fig:cartoon}(a), and (2) circuits that involve frequent measurement and reset of the ancilla, as shown in \autoref{fig:cartoon}(d).  The latter approach is more technically involved. 
In the following, we discuss the construction of the corresponding estimator and why it can be used to effectively realise the composite LCU with many segments.

\subsection{The symmetry }

Suppose that in the $k$th segment, $U_{i_k}$ and $V_{j_k}$ are sampled randomly and independently according to the probability distribution defined in \autoref{eq:RTE_LCU}. The notation $V_{j_k}$ is used to help the reader track its role explicitly, although it is sampled from the same LCU formula and in the same manner as $U_{i_k}$. Denote the ensembles of indices corresponding to the sampled unitaries over the $k$ segments as $\mathbf{i} := \{i_k\}_{k=1}^{\nu}$ and $\mathbf{j} :=  \{j_k\}_{k=1}^{\nu}$, respectively.

To start with, let us consider the case of $\nu = 1$.
Suppose that the unitary has an LCU form as $U = \mu \sum_i \Pr(i) U_i$. Let us denote $U^{\dagger} = \mu \sum_j \Pr(j) V_j^{\dagger} $. 
It is instructive, though slightly redundant, to write the expression in full to clearly illustrate the structure
$$ U\rho U^{\dagger} = \mu ( \sum_i \Pr(i) U_i) \rho (\sum_j \Pr(j) V_j^{\dagger}), $$
which can be equivalently written as
$$ (U^{\dagger}) ^{\dagger} \rho U^{\dagger} = \mu ( \sum_j \Pr(j) V_j) \rho (\sum_i \Pr(i) U_j^{\dagger}). $$

Now, given the sampled $U_i$ and $V_j$, by using this symmetrisation, we can see that the ensemble average takes the form
$$ \mathbb{E}_{i,j} \frac{\mu^2}{2} ( U_i \rho V_j^{\dagger} + V_j \rho U_i^{\dagger}   ) =  U\rho U^{\dagger}. $$

Then let us consider the case with two segments
$ 
 U_2 U_1\rho U_1^{\dagger} U_2^{\dagger}.
$
For simplicity, we consider the case where $U_1 = U_2 = U$; the more general case follows analogously. The symmetry is manifested as follows:
\begin{equation}
\label{eq:sym_state}
  \mathscr{S}_{i_1, j_1, i_2, j_2}(\rho) :=\frac{1}{4} (  U_{i_2} U_{i_1} \rho V_{j_1}^{\dagger} V_{j_2}^{\dagger} + U_{i_2} V_{j_1} \rho U_{i_1}^{\dagger} V_{j_2}^{\dagger} 
+ V_{j_2} V_{j_1} \rho U_{i_1}^{\dagger} U_{i_2}^{\dagger} 
+  V_{j_2} U_{i_1} \rho V_{j_1}^{\dagger} U_{i_2}^{\dagger}
).  
\end{equation}   
Here, we have defined the operation that generates the symmetrised, paired state as $\mathscr{S}_{i_1, j_1, \ldots, i_k, j_k}(\cdot)$, which produces a state containing all possible symmetric configurations.  The symmetry here refers to the pairing of $(i_k, j_k)$ for each $k$, while indices $i_k$ and $i{k'}$ (for $k \neq k'$) remain independent.
Note that $\mathscr{S}$ is a linear operation. We shall use its linearity to prove Proposition 2.

It is easy to check that 
\begin{equation}
    \mathbb{E}_{i,j} \mu_T^2  \mathscr{S}_{i_1, j_1, i_2, j_2}(\rho) =  U\rho U^{\dagger}.
\end{equation}

Back to the output state generated by the sampled unitaries $U_i$ and $V_j$, by taking the average over $a$, for $b = 0$,  we have
\begin{equation}
  \mathbb{E }_a (-1)^a  \sigma_{a, b = 0} =  \frac{1}{2} (U_i  \rho V_j^{\dagger} + V_j  \rho U_i^{\dagger} ) = \mathscr{S}_{i_1, j_1}(\rho).
\end{equation}
In the next section, we will extend it to multiple segments.

% A first observation is that 
% \begin{equation}
%     \math
% \end{equation}

\subsection{Estimator in the case of frequent measurement and reset}

In the main text, we have shown that averaging over $\mathbf{i}$ and $\mathbf{j}$ yields an unbiased realisation of the composite LCU acting on an arbitrary state, c.f. Proposition 2. Here, we delve deeper into the case of the frequent measurement and reset circuit by examining what each individual estimator will be.

Recall that the process of the $k$th segment is denoted as $\mathcal{E}_{a_{k}, b_{k}} $. 
% With the input state $\tr(\sigma) = 1$, the resulting state is 
% \begin{equation}
%     \mathcal{E}_{a_k, b_k} (\sigma) := \frac{K_{a_{k}, b_{k}}  \sigma  K_{a_{k} |b_{k}}^{\dagger}}{  \tr(K_{a_{k}, b_{k}} \sigma  K_{a_{k} |b_{k}}^{\dagger} )}. 
% \end{equation}
% Here,
% $ \mathcal{E}_{a_k, b_k}$ is a non-linear operation that depends on the outcomes $a_k, b_k$.  $K_{a_{k}, b_{k}}$ is a generalised operation (we shall derive its explicit form later). \sun{It is easy to see that 
% $$\sum_{a_{k}, b_{k}}  K_{a_{k} |b_{k}}  K_{a_{k} |b_{k}}^{\dagger} \neq I. $$}
The output state in each segment is defined iteratively
\begin{equation}
    \sigma^{(k)} = \mathcal{E}_{a_k, b_k}(\sigma^{(k-1)}) / \tr( \mathcal{E}_{a_k, b_k}(\sigma^{(k-1)})).
\end{equation}
Here $ \sigma^{(k)}$ is an abbreviated notation of the output state as it contains many configurations depending on the historic outcomes $\{ ( a_k, b_k ) \}_{k}$.

% The normalised state after $\nu$ segments of applying operations and resetting is denoted as \begin{equation}
%     \rho^{(\nu)} := \mathcal{E}_{a_{\nu}, b_{\nu}} \circ \mathcal{E}_{a_{\nu-1}, b_{\nu-1}} \circ ... \circ \mathcal{E}_{a_{1}, b_{1}} (\rho) 
% \end{equation}
% where $\mathcal{E}_{a_{\nu}, b_{\nu}} :=  K_{a_{\nu} |b_{\nu}} (\cdot) K_{a_{\nu} |b_{\nu}}^{\dagger}$ and 
% The normalised state can be expressed as $\sigma^{(\nu)} =  \rho^{(\nu)} / \tr ( \rho^{(\nu)} ) $.

As discussed in the main text, to realise composite LCU, we only need to consider the case where either $b_{k} = 0$ or $b_{k} = 1$. 
A unified definition of the estimator is
\begin{equation}
   v ^{(\nu)} := i^{\sum_{k=1}^{\nu} b_{k}} (-1)^{\sum_{k=1}^{\nu} a_k}  \sigma^{(\nu)}.
\end{equation}
% Denote $\mathbf{a} = \{ a_k\}_{k=1}^{\nu}$ and $\mathbf{b} = \{ b_k\}_{k=1}^{\nu}$.
In both cases, there is no randomness in $b_k$.
We shall prove that this is an unbiased estimator by the iterative method.

Let us consider the generalised quantum operations in the last step.
Let us denote the input normalised state before the application of $\mathcal{E}_{a_{\nu}, b_{\nu}}$ by $\sigma$ with $\tr(\sigma) = 1$. Then, after the application $\mathcal{E}_{a_{\nu}, b_{\nu}}$, the resulting normalised  state becomes
\begin{equation}
     \sigma_{a_{\nu}, b_{\nu}} =  \frac{1}{  \Pr(a_{\nu}|b_{\nu})}  \frac{1}{4} ( (-i)^{b_{\nu}}  (-1)^{a_{\nu}} U_{\nu} + V_{\nu} ) \rho (  i^{b_{\nu}} (-1)^{a_{\nu}} U_{\nu}^{\dagger} + V_{\nu}^{\dagger} ).
\end{equation}

In the last step, by taking the average over $a_{\nu} $ given $b_k = 0$, we have
\begin{equation}
\label{eq:QInstrument_avr}
\begin{aligned}
    \mathbb{E}_{a_{\nu}}    (-1)^{  a_{\nu}}     \sigma_{a_{\nu} } &= \sum_{a_{\nu} } \Pr(a_{\nu}| b_{\nu}) K_{a_{\nu} |b_{\nu}} \sigma^{(\nu-1)} K_{a_{\nu} |b_{\nu}}^{\dagger} /\tr(K_{a_{\nu} | b_{\nu}} \sigma^{(\nu-1)} K_{a_{\nu} |b_{\nu}}^{\dagger})  \\
    &=   \frac{1}{2} \left(  U_{\nu} \sigma V_{\nu}^{\dagger}   + V_{\nu} \sigma U_{\nu}^{\dagger} \right).
\end{aligned}
\end{equation}

Therefore, we have 
\begin{equation}
    \begin{aligned}
      \mathbb{E}_{a_{\nu}, ..., a_1} v^{(\nu)}  & =  \mathbb{E}_{a_{\nu}, ..., a_1}  (-1)^{\sum_{k=1}^{\nu} a_k}  \sigma^{(\nu)}   \\
        &=         \mathbb{E}_{a_{\nu-1},  ..., a_1}  (-1)^{\sum_{k=1}^{\nu-1} a_k} \mathbb{E}_{a_{\nu}} K_{a_{\nu} |b_{\nu}} \sigma^{(\nu-1)} K_{a_{\nu} |b_{\nu}}^{\dagger} / \tr(K_{a_{\nu} | b_{\nu}} \sigma^{(\nu-1)} K_{a_{\nu} |b_{\nu}}^{\dagger})\\
        &=         \mathbb{E}_{a_{\nu-1},  ..., a_1}  (-1)^{\sum_{k=1}^{\nu-1} a_k}   \sum_{a_{\nu} } \Pr(a_{\nu}| b_{\nu}) K_{a_{\nu} |b_{\nu}} \sigma^{(\nu-1)} K_{a_{\nu} |b_{\nu}}^{\dagger} /\tr(K_{a_{\nu} | b_{\nu}} \sigma^{(\nu-1)} K_{a_{\nu} |b_{\nu}}^{\dagger}) \\
        &=         \mathbb{E}_{a_{\nu-1},  ..., a_1}  (-1)^{\sum_{k=1}^{\nu-1} a_k}   \cdot \mathscr{S}_{i_\nu, j_\nu}(\sigma^{(\nu-1)} ) \\
        &=       \mathscr{S}_{i_\nu, j_\nu}   \left(  \mathbb{E}_{a_{\nu-1},  ..., a_1}  (-1)^{\sum_{k=1}^{\nu-1} a_k}   \sigma^{(\nu-1)} \right)    \\ 
         &=    \mathscr{S}_{i_1, j_1, \ldots, i_k, j_k}(\rho)
    \end{aligned}
\end{equation}

Here the average is only taken over $a_k$.
The key is to iteratively use \autoref{eq:QInstrument_avr} which holds when we are dealing with the currently last step and the linearity of $\mathscr{S}_{i_1, j_1, \ldots, i_k, j_k}(\rho)$ is used.

Proposition 2 can be easily proven by using
\begin{equation}
    \mathbb{E}_{i_1, j_1, \ldots, i_k, j_k} \mu_T^2 \mathscr{S}_{i_1, j_1, \ldots, i_k, j_k}(\rho) = U^\nu \rho (U^\nu)^{\dagger}.
\end{equation}

For the case with $b_k = 1$, we have
\begin{equation}
\label{eq:QInstrument_avr}
\begin{aligned}
    \mathbb{E}_{a_{\nu}}    i^{  b_{\nu}} (-1)^{a_{\nu}}     \sigma_{a_{\nu} }     &=   \frac{1}{2} \left(  U_{\nu} \sigma V_{\nu}^{\dagger}   - V_{\nu} \sigma U_{\nu}^{\dagger} \right).
\end{aligned}
\end{equation}
This construction could also be used to generate another type of symmetrised state  in analogue to $\mathscr{S}_{i_1, j_1, \ldots, i_k, j_k}$, which we leave for interested readers to explore.

\section{Variance of the estimator}

% Define $\hat{o}_m :=  \tr( O_m \hat{v}_{a,b}) $ for observable $O_m$.
% It is easy to check that
% \begin{equation}
%    \mathbb{E}_{R, a, \vec{z}}\hat{o}_m =  \tr( O U_m \rho V^{\dagger}) 
% \end{equation}

% just  need to compute 
% \begin{equation}
%     \tr ( \mathcal{M}^{-1}(O)   R^\dagger |\vec{z}\rangle \langle \vec{z}| R  )
% \end{equation}
% due to self-duality of $\mathcal{M}$.
% The advantage is that the inverted channel, whose form depends on the random ensemble, acts only on the observable, not on the state.

For observable estimation $O$ by the circuit \autoref{fig:cartoon}(a,b), it is easy to check that
\begin{equation}
   \mathbb{E}_{a,b, R, \vec{z}} (\hat{o}) = \mathbb{E}_{a,b,R,  \vec{z}} \tr (O   \hat{v}_{a, b} ) =  \tr( O U \rho V^{\dagger}). 
\end{equation}
Below, we discuss the variance of $\hat{o}$, which is related to the sampling cost.
\begin{equation}
\begin{aligned}
\mathbb{E}_{a,b, R, \vec{z}}(\hat{o}^2) & \leq \sum_{b = 0,1} \mathbb{E}_{ a  = 0, 1} \Pr(a) \, \mathbb{E}_{R \in \mathcal{C}} \sum_{\vec{z} \in \mathbb{Z}_2^n} \Pr(\vec{z}| a) \, \operatorname{Tr} \left[ O \mathcal{M}^{-1}( R^\dagger \ket{\vec{z}^{(b)}} \bra{\vec{z}^{(b)}} R) \right]^2 \\
& = \sum_{b = 0,1} \mathbb{E}_{ a  = 0, 1} \Pr(a) \, \mathbb{E}_{R \in \mathcal{C}} \sum_{\vec{z} \in \mathbb{Z}_2^n} \Pr(\vec{z}| a) \, \operatorname{Tr} \left[ \mathcal{M}^{-1}(O) R^\dagger \ket{\vec{z}^{(b)}} \bra{\vec{z}^{(b)}} R \right]^2 \\
&= \sum_{b = 0,1}  \, \mathbb{E}_{R \in \mathcal{C}} \sum_{\vec{z} \in \mathbb{Z}_2^n} \bra{\vec{z}} R  ( \mathbb{E}_{ a  = 0, 1} \Pr(a) \sigma_{a,b} ) R^\dagger \ket{\vec{z}}  \operatorname{Tr} \left[ \mathcal{M}^{-1}(O) R^\dagger \ket{\vec{z}^{(b)}} \bra{\vec{z}^{(b)}} R \right]^2 \\
&= 2 \sum_{R \in \mathcal{C}} \sum_{\vec{z} \in \mathbb{Z}_2^n} \bra{\vec{z}} R \left( U \rho U^\dagger + V \rho V^\dagger \right) R^\dagger \ket{\vec{z}} \, \operatorname{Tr} \left[ \mathcal{M}^{-1}(O) R^\dagger \ket{\vec{z}^{(b)}} \bra{\vec{z}^{(b)}} R \right]^2 \\
&\leq  2 \|O\|_{\text{shadow}}^2.
\end{aligned}
\end{equation}
Here we have used $(x + i y)^2 \leq |x|^2 + |y|^2$ in the first inequality. In  the second equation, we use the self-duality of $\mathcal{M}$, which is
\begin{equation}
    \tr (\mathcal{M}^{-1}(O) \rho ) =   \tr (\mathcal{M}^{-1}( \rho)  O)
\end{equation}
for any $\rho \in \mathcal{H}(2^n)$.
We have also used $$\mathbb{E}_{a  = 0, 1} \Pr(a) \sigma_{a,b} = U \rho U^{\dagger} + V \rho V^{\dagger}.$$

% Note that we can define
% \begin{equation}
%      \tr ( \mathcal{M}^{-1}(O)   \hat{v}_{a,b} ) =  \sum_b \tr (O  i^b (-1)^a \mathcal{M}^{-1} ( \hat{\rho}_{a, b} (R, \vec{z}))).
% \end{equation}
% The advantage is that the inverted channel, whose form depends on the random ensemble, acts only on the observable, not on the state.

The variance of $\hat{o}$ is thus bounded by $\Var (\hat{o} ) \leq  2 \|O\|_{\text{shadow}}^2 - \tr ( O_m U \rho V^{\dagger})^2$.

\section{Shadow estimation for states generated by circuits with frequent measurement and reset}

With the bitstring $\vec{z}^{(b)}$ obtained from randomised measurements, we can get an unbiased snapshot of the state $ \rho^{(\nu)}$ as
\begin{equation}
    \hat{\rho}^{(\nu)}(R, \vec{z})
=  \mathcal{M}^{-1} \left( R^\dagger |\vec{z}\rangle \langle \vec{z}| R \right)
% \label{eq:shadow_estimator_v_a}
\end{equation} with $\mathbb{E}_{R, \vec{z}} \left( \hat{\rho}^{(\nu)} (R, \vec{z}) \right) = \sigma^{(\nu)}$,  where $\mathcal{M}^{-1}  $ is a invertible linear map determined by $\mathcal{C}$.

Recall that for the three applications considered in this work, only one circuit instance is needed, i.e. either $\mathbf{a} = 0$ or $\mathbf{b} = 0$ where $\mathbf{a} = \{ a_k\}_{k=1}^{\nu}$ and $\mathbf{b} = \{ b_k\}_{k=1}^{\nu}$.
For $\mathbf{b} = 0$, we can define
\begin{equation}
    \hat{v}^{(\nu)}_{\mathbf{a}, \mathbf{b} = 0}(R, \vec{z}) = \mu_T^2   (-1)^{\sum_{k=1}^{\nu} a_k}   \hat{\rho}^{(\nu)}(R, \vec{z}).
\end{equation}

One can show that $$   \mathbb{E}_{\mathbf{a}, \mathbf{b}, R, \vec{z}}  \hat{v}^{(\nu)}_{\mathbf{a},\mathbf{b} = 0}(R, \vec{z}) =   \mathscr{S}_{i_1, j_1, \ldots, i_k, j_k}(\rho)  $$
And we have  $$  \mu_T^2 \mathbb{E}_{\mathbf{i},\mathbf{j}}\mathbb{E}_{\mathbf{a}, \mathbf{b} = 0, R, \vec{z}}  \hat{v}^{(\nu)}_{\mathbf{a}, \mathbf{b}}(R, \vec{z}) =   U^{\nu} \rho   (U^{\nu})^{\dagger}  $$
where the normalisation factor in the composite LCU form is $\mu_T$ as defined in this main text.
% The variance can be similarly derived. 
For $\mathbf{b} = 1$, the explicit formula can be similarly derived based on another symmetrised operation in analogue to $\mathscr{S}_{i_1, j_1, \ldots, i_k, j_k}$.

% \bibliographystyle{unsrt}
% \bibliography{resource}

\end{document}